\documentstyle[sprocl]{article}

\def\Journal#1#2#3#4{{#1} {\bf #2}, #3 (#4)}

\def\PLB{{\em Phys. Lett.}  B}
\def\PRL{\em Phys. Rev. Lett.}
\def\PRD{{\em Phys. Rev.} D}

\def\be{\begin{equation}}
\def\ee{\end{equation}}
\def\bea{\begin{eqnarray}}
\def\eea{\end{eqnarray}}
\newcommand{\br} {{\bf r}}
\newcommand{\brd} {{\dot{\bf r}}}
\newcommand{\bR} {{\bf R}}
\newcommand{\bRd} {{\dot{\bf R}}}
\newcommand{\bsigma} {{\mbox{\boldmath ${{\bf \sigma}}$}}}
\newcommand{\brho} {{\mbox{\boldmath ${{\bf \rho}}$}}}
\newcommand{\blambda} {{\mbox{\boldmath ${{\bf \lambda}}$}}}
\newcommand{\bsigmad} {{\mbox{\boldmath ${{\dot{\bf \sigma}}}$}}}
\newcommand{\brhod} {{\mbox{\boldmath ${{\dot{\bf \rho}}}$}}}
\newcommand{\blambdad} {{\mbox{\boldmath ${{\dot{\bf \lambda}}}$}}}
\newcommand{\omr} {{\omega_{\rho}}}
\newcommand{\oml} {{\omega_{\lambda}}}
\newcommand{\oms} {{\omega_{\sigma}}}
\newcommand{\omsd} {{\omega_{\sigma}^{'}}}
\newcommand{\barr}[1]{\begin{array}{#1}}
\newcommand{\earr}{\end{array}}
\newcommand{\beqna}{\begin{eqnarray}}
\newcommand{\eeqna}{\end{eqnarray}}

\begin{document}

\title{\begin{flushright} 
\small{nucl-th/0004053}
\\ \small{LA-UR-00-2054}
 \end{flushright} 
\vspace{0.8cm}  
HYBRID BARYON SIGNATURES}

\author{P. R. PAGE\footnote{
Talk at ``The Physics of Excited Nucleons'' (NSTAR2000), Newport News, VA,
16--19 Feb. 2000.
}}

\address{Theoretical Division, MS-B283, Los Alamos National 
Laboratory, \\ Los Alamos, NM 87545, USA\\E-mail: prp@lanl.gov} 

\maketitle\abstracts{We discuss whether a low--lying
{\it hybrid baryon} should be 
defined as a three quark -- gluon bound state or as three quarks moving on 
an excited adiabatic potential. We show that the latter definition becomes
exact, {\it not} only for very heavy quarks, but also for specific 
dynamics. We review the literature on the signatures of hybrid baryons,
with specific reference to strong hadronic decays, electromagnetic couplings,
diffractive production and production in $\psi$ decay.}

\section{What is a hybrid baryon?}

Historically a low--lying hybrid baryon was defined as a three quark -- gluon 
composite. However, from the viewpoint of the Lagrangian of 
Quantum Chromodymanics (QCD) this definition is non--sensical.
This is because gluons are massless, and hence there is no reason not
to define a hybrid baryon, for example, as a three quark -- two gluon 
composite. Neither is one possibility distinguishable from the other,
since strong interactions mix the possibilities. The place where this
definition of a hybrid baryon is most
useful is in large $Q^2$ deep inelastic scattering, where a Fock state
expansion of a state can rigorously be defined, and one can at least
talk about the three quark -- gluon component of such a state.\cite{carlson}
However, in other situations the definition becomes
perilous. A case in point is recent work on large $N_c$ hybrid
baryons, where their properties depend critically on the fact that the
gluon is in colour octet, and hence the three quarks in colour octet,
so that the entire state is colour singlet.\cite{pirjol}
The bag model circumvents the objections raised against this definition,
since gluons become massive due to their confinement inside the bag.\cite{bag}

More recently, a low--lying hybrid baryon was defined as three quarks moving
on the low--lying
excited adiabatic potential.\cite{paton85} From the viewpoint of
QCD this can be a perfectly sensible definition. One can always evaluate
the energy of a system of three fixed quarks as a function of the three
quark positions, called the adiabatic potential. A calculation
along these lines has been performed in flux--tube models\cite{capstick} 
and a first attempt has been made in lattice QCD.\cite{bali}
The three quarks are then allowed to move in a three--body equation,
typically a non--relativistic
Schr\"{o}dinger equation. 
Treating a three quark system via this two step process
is called the adiabatic or Born--Oppenheimer approximation. 
Note that it is not appropriate to talk about the case where the quarks are
actually infinitely heavy, because of the lack of kinetic energy. 
The adiabatic approximation is expected to become
exact if the quark masses are much greater than the scale of the
strong interactions $\Lambda_{QCD}$. The criterion for the 
validity of the adiabatic approximation is that the slow degrees of freedom
(the quarks) should move much slower than the fast
degrees of freedom (the gluons). It is possible to argue that for
conventional baryons the relative velocities of quarks behave like the strong
coupling constant $\alpha_S$ as the
quarks become heavier.\cite{isgur} Because of the asymptotic freedom of
QCD, the quark relative velocities go to zero, ensuring the validity of the
adiabatic approximation. In fact, this is the basis for the NRQCD
expansion. However, since $\alpha_S$ goes to zero only logarithmically, 
one may need 
quarks heavier than the bottom quark for the adiabatic approximation
to be valid. Depending on the shape of the adiabatic
potential the possibility of an NRQCD expansion for {\it hybrid} 
baryons,\cite{petrov} and hence the 
validity of an adiabatic approximation, may be in jeopardy.

Here we point out for the first time 
that for specific dynamics, the adiabatic approximation
can be {\it exact}, even for light quarks, if one redefines the
adiabatic potential suitably. We call this the ``redefined adiabatic
approximation'', which employs a ``redefined adiabatic
potential''. It was noted in a flux--tube model that
``For light quarks almost all corrections may be incorporated into a 
redefinition of the potentials. Mixing between [new] potentials 
is of the order 
of 1\%''.\cite{merlin} 
We develop the following technique for obtaining the
redefined potential. The quark positions are still fixed
relative to each other, but the quark and gluon positions are not defined 
relative to the quark positions, but relative to the centre of mass
of the quarks and the gluons, as recently pioneered.\cite{capstick}
For dynamics where the redefined adiabatic approximation is exact
one can rigorously define a low--lying hybrid baryon as three quarks moving
in the low--lying redefined excited adiabatic potential.

\section{Redefined adiabatic approximation} 

Consider the non--relativistic hamiltonian for three quarks 
at positions $\br_1,\br_2,\br_3$ and a junction at position $\br_4$.
The junction represents the gluons. The classical
hamiltonian with a simple harmonic oscillator potential is

\begin{equation} \label{ham}
H = \frac{1}{2}M(\brd^2_1+\brd^2_2+\brd^2_3) + \frac{1}{2} m \brd^2_4
+\frac{1}{2} k ((\br_1-\br_4)^2+(\br_2-\br_4)^2+(\br_3-\br_4)^2)
\end{equation}
where $M,m$ and $k$ are the mass of the quarks, the mass of the junction
and coupling constant respectively. The first two and the last terms are the
kinetic and potential energy terms respectively. 

\subsection{Exact solution}

Making the variable 
transformations

\[
\brho = \frac{1}{\sqrt{2}}(\br_1-\br_2) \hspace{1cm}
\blambda = \frac{1}{\sqrt{6}}(\br_1+\br_2-2\br_3)
\]

\vspace{-.5cm}
\begin{equation} \label{tran}
\bsigma = \frac{2}{\sqrt{3}}\frac{m}{m+3M}(\br_1+\br_2+\br_3-3 \br_3)
\end{equation}
one obtains the diagonal hamiltonian

\begin{equation} \label{hame}
H = \frac{1}{2} M (\brhod^2+\blambdad^2+\frac{1}{4}(1+3\frac{M}{m})\bsigmad^2) + \frac{1}{2} k\; (\brho^2+\blambda^2+\frac{1}{4}(1+3\frac{M}{m})^2\bsigma^2)
\end{equation}
when working in the overall centre of mass frame. The hamiltonian  
describes three independent simple harmonic oscillators so that the
energies solving the (quantized) Schr\"{o}dinger equation,  are
\begin{equation} \label{ene}
E = (n_\rho+\frac{3}{2})\omr+(n_\lambda+\frac{3}{2})\oml+(n_\sigma+\frac{3}{2})\oms
\end{equation}
where e.g. $n_\rho = n_\rho^x +n_\rho^y +n_\rho^z$, with $n_\rho^x,\; 
n_\rho^y$ and $n_\rho^z$ the degrees of excitation in the three space
directions. The vibrational frequencies are

\begin{equation}
\omr^2 = \oml^2 = \frac{k}{M}\hspace{1.5cm} 
\oms^2 = (1+3\frac{M}{m}) \frac{k}{M}
\end{equation}
The wave functions solving the Schr\"{o}dinger equation can be denoted by

\begin{equation} \label{wave}
\psi_{n_\rho}(\brho)\;\psi_{n_\lambda}(\blambda)\;\psi_{n_\sigma}(\bsigma) 
\end{equation}
where e.g. $n_\rho$ denotes the set $\{n_\rho^x,n_\rho^y,n_\rho^z\}$.

One can verify that the quark relative velocities 
$\sim (\frac{k}{M m^2})^{\frac{1}{4}}$, so that they go to zero
if $M \gg \frac{k}{m^2}$. The criterion for the validity of the 
adiabatic approximation is hence satisfied for large quark masses.

\subsection{Adiabatic solution}

Fix the quark positions so that $\brd_1=\brd_2=\brd_3=0$. 
The potential energy depends only on the differences between positions,
and is hence unaffected by fixing the quark positions.
The kinetic energy in Eq. \ref{ham} is $\frac{1}{2}m\brd_4^2$, and
using $2\sqrt{3}\brd_4=-(1+3\frac{M}{m})\bsigmad$ from Eq. \ref{tran},
the hamiltonian (\ref{ham}) is

\begin{equation}
H_j = \frac{1}{2} \frac{m}{12} (1+3\frac{M}{m})^2 \bsigmad^2 +
\frac{1}{2} k\; (\brho^2+\blambda^2+\frac{1}{4}(1+3\frac{M}{m})^2\bsigma^2)
\end{equation}
Only the variable $\bsigma$ is dynamical. The energies (adiabatic potentials)
are

\begin{equation}\label{adpot}
E_j = (n_\sigma+\frac{3}{2})\omsd + \frac{1}{2} k (\brho^2+\blambda^2)
\hspace{1cm} \omsd^{2} = 3 \frac{k}{m}
\end{equation}
Now allow the quarks to move in their centre of mass frame, 
so that the quark motion hamiltonian is

\begin{eqnarray}
H_q &  = & \frac{1}{2}M(\brd^2_1+\brd^2_2+\brd^2_3)  + E_j\nonumber \\ 
 &   =  & \frac{1}{2}M(\brhod^2+\blambdad^2) + 
(n_\sigma+\frac{3}{2})\omsd + \frac{1}{2} k (\brho^2+\blambda^2)
\end{eqnarray}
The energies are

\begin{equation}
E_q = (n_\rho+\frac{3}{2})\omr+(n_\lambda+\frac{3}{2})\oml+(n_\sigma+\frac{3}{2})\omsd
\end{equation}
It is easy to see that these adiabatic approximation energies only
agree with the exact energies (\ref{ene}) when $\frac{M}{m} \gg 1$, i.e.
when the quarks are much heavier than the gluons. This is in accord
with one's na\"{\i}ve expectation for the validity of the adiabatic
approximation.

\subsection{Redefined adiabatic solution}

We follow the same procedure as for the adiabatic solution, with one
critical change. The quark positions are still fixed will respect to
each other, but all positions are now defined with respect to the
overall centre of mass, before we allow the quarks to move. Define
the overall centre of mass as

\begin{equation}
\bR = \frac{M(\br_1+\br_2+\br_3) + m\br_4}{m+3M}\hspace{.3cm} \Rightarrow
\hspace{.3cm}\bRd = \frac{m}{m+3M} \brd_4 = -\frac{1}{2\sqrt{3}} \bsigmad 
\end{equation}
Define new coordinates which are the
positions of each particle with respect to the overall centre of
mass. The time derivatives of these coordinates are

\begin{equation}
\brd_1^{'} = \brd_2^{'} = \brd_3^{'} = -\bRd  =
\frac{1}{2\sqrt{3}} \bsigmad \hspace{1cm}
\brd_4^{'} = \brd_4 -\bRd = -\frac{\sqrt{3}}{2}\frac{M}{m}\bsigmad  
\end{equation}
The kinetic energy in terms of the new coordinates is

\begin{equation}
\frac{1}{2}M(\brd^{'\; 2}_1+\brd^{'\; 2}_2+\brd^{'\; 2}_3) + 
\frac{1}{2} m \brd^{'\; 2}_4 = \frac{1}{2}
\frac{M}{4} (1+3\frac{M}{m}) \bsigmad^2
\end{equation}
This kinetic energy combined with the potential in Eq. \ref{ham} is

\begin{equation}
H_j = \frac{1}{2}\frac{M}{4} (1+3\frac{M}{m}) \bsigmad^2 +
\frac{1}{2} k\; (\brho^2+\blambda^2+\frac{1}{4}(1+3\frac{M}{m})^2\bsigma^2)
\end{equation}
We used the fact that the potential energy depends only on the differences 
between positions. The energies (redefined adiabatic potentials)
are

\begin{equation}\label{adrpot}
E_j = (n_\sigma+\frac{3}{2})\oms + \frac{1}{2} k (\brho^2+\blambda^2)
\end{equation}
and the junction wave functions $\psi_{n_\sigma}(\bsigma)$.
Allowing the quarks to move, the quark motion hamiltonian is

\begin{equation}
H_q   = \frac{1}{2}M(\brhod^2+\blambdad^2) + 
(n_\sigma+\frac{3}{2})\oms + \frac{1}{2} k (\brho^2+\blambda^2)
\end{equation}
The energies are

\begin{equation}
E_q = (n_\rho+\frac{3}{2})\omr+(n_\lambda+\frac{3}{2})\oml+(n_\sigma+\frac{3}{2})\oms
\end{equation}
which are identical to the exact solution (\ref{ene}). The quark 
wave functions are $\psi_{n_\rho}(\brho)\;\psi_{n_\lambda}(\blambda)$.
In order to obtain the full wave functions of the system, we take the 
direct product of the  quark and junction wave functions, i.e.
$\psi_{n_\rho}(\brho)\;\psi_{n_\lambda}(\blambda)\;\psi_{n_\sigma}(\bsigma)$.
These are identical to the wave functions of the 
exact solution (\ref{wave}). 

In conclusion, the solution obtained via the redefined adiabatic
approximation is {\it exact}.

We shall now show how the specific dynamics (\ref{ham}) provide a toy model
for the quark model, in the sense that many of the features needed for the
validity of quark model, are {\it exact} in the toy model. Quark models
enable the dynamics of quarks, while freezing out the dynamics of gluons.
In the toy model this corresponds to claiming that gluons are in the 
same wave function for all conventional baryons. But this is manifestly
the case in the redefined adiabatic approximation. 
The redefined adiabatic potential (\ref{adrpot}) is explicitly
dependent on quark mass. This corresponds to quark models: the Coulomb
interaction is usually postulated to have a coupling that depends on 
quark mass. On the other hand, the adiabatic potential (\ref{adpot})
is not dependent on quark mass. 

The wave functions obtained from the adiabatic approximation differ
from those in the redefined adiabatic approximation only for the gluons.
When one freezes the dynamics of gluons, as in the quark model, one 
does not notice the difference between these two approximations.
Even the functional forms of the gluon wave functions are the same: 
the wave functions only differ in the dimensionful scale they contain.
Thus the forms can be used interchangeably. 

When a process is studied that involves more than one redefined
adiabatic potential, e.g. hybrid baryon decay to a conventional baryon 
and meson, care has to be taken to use the redefined adiabatic gluon 
wave functions instead of the adiabatic ones. The former gluon wave functions
are dependent on $\oms$, which is itself dependent on quark mass.

\section{Where can one search for hybrid baryons?}

We first outline the results of a recent flux--tube model 
calculation.\cite{capstick} 
The flavour, non--relativistic spin $S$ and $J^{P}$ of the seven low--lying 
hybrid baryons \\ are 
$(N,\Delta)^{2S+1}J^P =
N^2 {\frac{1}{2}}^+, \; N^2 {\frac{3}{2}}^+, \; \Delta^4 {\frac{1}{2}}^+, 
\; \Delta^4 {\frac{3}{2}}^+, \; \Delta^4 {\frac{5}{2}}^+$, 
where the first two states double.  The bag model has the same number of
states. The pair $N^2 {\frac{1}{2}}^+, \; N^2 {\frac{3}{2}}^+$
has the same quantum numbers as in the flux-tube model.\cite{bag}
The other five states in the bag model are
$ \Delta^2 {\frac{1}{2}}^+, \; \Delta^2 {\frac{3}{2}}^+, \; 
N^4 {\frac{1}{2}}^+, \; N^4 {\frac{3}{2}}^+, \; N^4 {\frac{5}{2}}^+$.
The state $N {\frac{1}{2}}^+$ was studied in QCD sum rules.\cite{sum}
The hyperfine interaction moves the $\Delta$ states upwards and the
$N$ states downwards.\cite{capstick} Hence there are four low--lying $N$ 
hybrid baryons with a mass of $1870\pm 100$ MeV. This is somewhat higher,
but sometimes within errors, of
 bag models and QCD sum rules which find a mass around 
1.5 GeV.\cite{bag,sum}
The wave function sizes are estimated as 
$\sqrt{\langle\rho^2\rangle} =  \sqrt{\langle\lambda^2\rangle} = 2.1$ 
(conventional baryons) and $2.5$ (hybrid baryons) GeV$^{-1}$. Hybrid
baryons are hence larger than conventional baryons.

The following techniques may enable the detection of hybrid baryons:

\noindent $\bullet$ {\it Overabundance of states}: 
This approach is troublesome, since
not even all conventional baryons in the appropriate mass region have
been discovered yet. However, hybrid baryon states are likely to be
discovered before all conventional baryons in a relevant mass region, and
hence need to be studied alongside conventional baryons.

\noindent $\bullet$  {\it Decays}: Except for a QCD sum rule 
motivated suggestion that
hybrid baryons should decay to $N\sigma$,\cite{li} no decay calculations
have been performed. However, decay of hybrid baryons to 
$N\rho$ and $N\omega$ is {\it a priori} interesting since it isolates
states in the correct mass region, without contamination from lower--lying
conventional baryons. Study of these decay modes will also yield 
information on photo-- and electroproduction in Hall B and C at 
Jefferson Lab, since the $\rho$ and $\omega$ couple to the photon via
vector meson dominance. The process 
$N\pi\rightarrow$ hybrid $\rightarrow N \omega$ can be studied at the
D--line at Crystal Ball E913.

\noindent $\bullet$ {\it Diffractive $\gamma N$ and $\pi N$ production:}
The detection of the hybrid meson candidate $\pi(1800)$ in diffractive
$\pi N$ collisions by VES\cite{ves} may indicate that hybrid mesons
are producted abundantly via meson--pomeron fusion. If this is the case,
one expects significant production of hybrid baryons via 
baryon--pomeron fusion, i.e. production in 
diffractive $\gamma N$ and $\pi N$ collisions.

\noindent $\bullet$ {\it Production in $\psi$ decays:} Na\"{\i}ve expectations
are that the gluon--rich environment of $\psi$ decays should lead to 
dominant production of glueballs, but also signifant production of
hybrid mesons and baryons. The large branching ratios\cite{pdg98}
$Br(\psi\rightarrow p\bar{p}{\omega},\;  p\bar{p}\eta^{'}) \sim
10^{-3}$  may contain hybrid baryons decaying to
$(p,\bar{p})\;\omega$ especially. Recently a 
$J^P=\frac{1}{2}^+$ $2\sigma$ peak at mass $1834^{+46}_{-55}$ MeV
was seen in $\Psi\rightarrow p\bar{p}\eta$.\cite{zou}

\noindent $\bullet$ {\it Electroproduction:} In the flux--tube model,
i.e. the adiabatic picture of a hybrid baryon, there
is the qualitative conclusion that
``$e p\rightarrow e X$ should produce ordinary $N^{\ast}$'s and hybrid 
baryons with comparable cross--sections''.\cite{bar}
However, the conclusions obtained from the three quark -- gluon picture
of a hybrid baryon is different. For large $Q^2$ electroproduction,
the $Q^2$ dependence of the amplitudes is summarized in Table 1.
Since the photon has both a transverse and 
longitudinal component, the amplitude for a conventional baryon 
is expected to dominate that of the hybrid baryon as $Q^2$ becomes 
large.\cite{carlson}
For small $Q^2$ the conclusion agrees with the large $Q^2$ result
for transverse photons, but is more dramatic for longitudinal photons:
the amplitude vanishes.\cite{volker}
It has accordingly been concluded that the ({radially } excited) conventional 
baryon is dominantly electroproduced, with the {hybrid} baryon subdominant 
to the resonances $S_{11}(1535), D_{13}(1520)$ and $\Delta$ as 
$Q^2$ increases.\cite{volker} The $Q^2$ dependence of the 
electroproduction of a resonance can be measured at 
Jefferson Lab Hall B and an
energy upgraded Jefferson Lab. A hybrid baryon is expected to behave
different from nearby conventional baryons as a function of $Q^2$. One
needs to perform partial wave analysis at different $Q^2$. For
large $Q^2$ cross--sections are small, which would make this way of
distinguishing conventional from hybrid baryons challenging.

\begin{table}[t]
\begin{center}
\caption{$Q^2$ dependence of amplitudes for the electroproduction of
conventional or hybrid baryons with transverse or longitudinal 
photons.\protect\cite{carlson}}
{\begin{tabular}{|c|c|c|}
\hline 
 & {Conventional}  & {Hybrid}  \\
\hline
{Transverse} & 1$/{Q^3}$ & $1/{Q^5}$ \\
{Longitudinal} & $1/{Q^4}$ & $1/{Q^4}$ \\
\hline 
\end{tabular} }
\end{center}
\end{table}


\section*{Acknowledgments}
The suggestion to study hybrid baryons with a simple harmonic oscillator
potential was made by Nathan Isgur and Simon Capstick. The manuscript
was carefully read by Noel Black.
This research is supported by the Department of Energy under contract
W-7405-ENG-36.

\end{document}